\documentstyle[aps,preprint,eqsecnum]{revtex}
\begin{document}
\title{ Cluster Distribution  in  Mean-Field
Percolation: Scaling and Universality}
\author{Joseph Rudnick and Paisan Nakmahachalasint}
\address{Department of Physics, UCLA, 405 Hilgard Ave. Los Angeles, California
90095-1547}
\author{George Gaspari}
\address{Department of Physics, University of California, Santa Cruz, Santa
Cruz, California 95064}
 \date{\today}
\maketitle
\begin{abstract}
The partition function of the finite $1+\epsilon$ state Potts model is
shown to yield a closed form for the distribution of clusters in the immediate
vicinity of the percolation transition. Various important properties of the
transition are manifest, including scaling behavior and the emergence of the
spanning cluster.
\end{abstract}
\section{Introduction}
One of the aspects of bond percolation that has captured the
imagination of researchers is the collection of scaling properties that a
percolating system exhibits in the vicinity of the transition at which the
spanning cluster emerges\cite{StaufferAharony}. These scaling properties
manifest
themselves in various correlation functions, which reveal the structure of
large
finite clusters, and through the moments of the  cluster size distribution
function. An adjunct of these scaling properties are quantities that are {\em
universal} at the percolation transition. Universality reflects the
insensitivity of behavior at and near the transition to details, and follows
from the dominating influence of long range correlations. The key feature
of the
percolation transition is the emergence of the spanning cluster, and the
concepts of scaling and universality are of great help in the description of
its characteristics in the vicinity of the transition.

Many of the scaling properties of the percolation transition, and all
information regarding the moments of the cluster distribution, are contained in
the generating function, defined in terms of the
distribution of cluster sizes as follows
\begin{equation} F(p,h) = \sum_m n^c_m
e^{-mh}    \label{generating}
\end{equation}
In the above, the quantity $n^c_m$ is the average number of clusters containing
$m$ sites, and $p$ quantifies the probability that a link between sites is
``active.'' An active link is called a bond. In the simplest versions of bond
percolation only those links that couple nearest neighbor sites will, with any
finite probability, be bonds.

In the standard version of percolation, only bonds connecting close-by sites
are allowed to be active. A version of percolation that lends itself to
exact analysis is the infinite range model investigated by Wu\cite{Wu} and,
as an
example of a random graph, by Erd\"{o}s and R\'{e}nyi\cite{ErdRen}. In this
model the probability that a bond exists between two sites is independent
of the
distance from one to the other. If there are $N$ sites, the probability that a
bond exists between any given pair is equal to $p/N$. In the ``thermodynamic
limit'' \mbox{$N \rightarrow \infty$} this quantity vanishes, but the effective
coordination number of each site diverges as $N$, and the net probability that
two sites are connected does not necessarily approach zero, or any other
trivial
value.

The infinite range model exhibits a percolation transition, in that the
probability, $P$, that that two arbitrarily chosen sites belong to the same
cluster, which is equal to zero in the thermodynamic limit when \mbox{$p\leq
1$}, takes on a finite value when \mbox{$p>1$}. The equation satisfied by the
quantity $P$ is
\begin{equation}
P = 1 - e^{-pP}    .  \label{equationofstate}
\end{equation}
The only non-negative solution to this equation is $P=0$ when $p\leq 1$, while
a non-zero solution exists when  $p > 1$. This latter solution saturates at
1 in
the limit \mbox{ $p = \infty$}. The quantity $P$ is also equal to the
fraction of
sites contained in the spanning cluster. This cluster contains  a finite
fraction of all the sites in the system find in the thermodynamic
limit.

The connection between the generating function and the statistical mechanics of
a particular model was established by Fortuin
and Kasteleyn\cite{FK}, who demonstrated the equivalence between the
percolation
generating function and the partition function of the $q \rightarrow 1$
limit of
the $q$-state Potts model\cite{Potts}. In particular, the generating
function as
given by Eq. (\ref{generating}) is, to within uninteresting factors, equal
to the
limiting ratio
\begin{equation}
\lim_{q \rightarrow 1} \frac{Z_q - Z_1}{q-1}   ,  \label{limit}
\end{equation}
where $Z_q$ is the partition function of the $q$-state Potts model. A number of
field-theoretical treatments of percolation are  based on the above
relation\cite{Harris}.

Given the generating function one can, in principle, determine the values of
the quantities $n^c_m$. This is because the sum in Eq. (\ref{generating}) has
the general form of a Laplace transform, and such transforms are readily
inverted. Given the dependence on $h$ of $F(p,Hh)$ one obtains the mean number
of clusters containing $n$ sites via \begin{equation}
n^c_m = \frac{1}{2\pi }\int_0^{2 \pi} F(p,ih) e^{-ihm} dh  \label{inverse}
\end{equation}

In this brief note we show that results previously obtained for the partition
function of the \mbox{$1 + \epsilon$-state} Potts model  with infinite range
interactions lead directly to the cluster distribution function for  the
infinite-range version of percolation described above. This distribution
function displays all the expected scaling properties, and in addition reveals
the precise way in which the spanning cluster emerges from the the ``sea'' of
 of clusters that  remain finite in the thermodynamic limit. As the model
investigated is  the ``mean field'' version of short-ranged
percolation, the results to be displayed for the cluster size distribution
represent  zeroth order approximations to the corresponding results relating to
the cluster size distribution function of the physically realizable, and
therefore physically relevant, short-ranged percolation\cite{us}.

\section{Scaling at the percolation transition}
The scaling laws that characterize the critical point have direct analogues in
the percolation transition. For instance, given the critical value of  $p$,
which we denote $p_c$, the percolation generating function in the immediate
vicinity of the transition takes on the following form
\begin{equation}
F\left(p_c(1+\Delta p), h\right) \rightarrow \left|\Delta p \right|^{w_1}
f\left( \left|\Delta p\right|^{-w_2}h, \frac{ \Delta p}{\left| \Delta p
\right|}
\right) . \label{scaling}
\end{equation}
The exponents $w_1$ and $w_2$  control the asymptotic behavior of
various aspects of the cluster size distribution. For example the $l^{\rm th}$
moment of the cluster distribution function, equal to the expectation valued of
the $l^{\rm th}$ moment of the cluster size $m$,  is given by
\begin{equation}
 \langle m^l \rangle =\frac{\sum_m m^l n^c_m}{\sum_n n^c_m} = \frac{\left.
(-1)^l\frac{d^l}{dh^l}F(p,h)\right|_{h=0}}{F(p,0)} ,   \label{moments1}
\end{equation}
as can be established by looking at Eq. (\ref{generating}).

Given the scaling form in Eq. (\ref{scaling}), the dependence of the
expectation value of $m^l$ on $\Delta p$ is
\begin{equation}
\langle m^l \rangle \propto \left|\Delta p\right|^{-lw_2} . \label{moments2}
\end{equation}
The constant of proportionality depends on sign of $\Delta p$. Eliminating the
denominator in Eqs. (\ref{moments1}), we have the following relations
\begin{equation}
\sum_nn^c_mm^l =  \left.\frac{d^l}{dh^l}F(p,h)\right|_{h=0}.
\end{equation}
when the power $l$ is equal to one, the right hand side is just the total
number of sites in the system, and this quantity is clearly idependent of the
clustering induced by the existence of bonds. This sum rule, which is violated
in the thermodynamic limit if when $h \rightarrow 0^-$ plays the role of the
symmetry that is violated when there is a symmetry-breaking transition.

Now, the scaling form in Eq. (\ref{scaling}), along with the  inversion
formula (\ref{inverse}) implies a cluster size distribution having  the
following scaling form
\begin{equation}
n^c_m = \left| \Delta p \right|^{w_1 + w_2} {\cal X} \left( m \left| \Delta p
\right|^{w_2}, \frac{ \Delta p }{\left| \Delta p \right|} \right) .
\label{scaling2}
\end{equation}
At the percolation transition, where \mbox{$\Delta p = 0$}, Equation
(\ref{scaling2}) implies a distribution function having the following form:
\begin{equation}
n^c_m \propto m^{-(w_1 + w_2)/w_2}      .  \label{criticalscaling}
\end{equation}
In the mean field limit, the two exponents $w_1$ and $w_2$ take on the
following values:
\begin{eqnarray}
w_1 &=& 3   \label{exp1} \\
w_2 &=& 2  \label{exp2}
\end{eqnarray}
Then, according to Eq. (\ref{criticalscaling}), $n^c_m \propto m^{-5/2}$.

When there is a finite number of sites in the system, the  cluster size
distribution  incorporates the number of sites, $N$, by taking on the more
general scaling form
\begin{equation}
n^c_m =\left| \Delta p \right|^{w_1 + w_2} {\cal X} \left( m \left|
\Delta p \right|^{w_2}, \frac{\left| \Delta p \right|^{-w_3}}{N},\frac{ \Delta
p }{\left| \Delta p \right|} \right) .  \label{scaling3}
\end{equation}
In the mean field limit, the exponent $w_3$ is equal to 3. Eq. (\ref{scaling3})
implies that the relation (\ref{criticalscaling}) applies at the percolation
transition until $m \propto N^{2/3}$.

\section{Cluster Size Distribution}
The distribution of cluster sizes in mean field percolation follows directly
from a result for the ratio in Eq. (\ref{limit}). This result was basd on an
analysis of the mean field version of the $q$ state Potts model in the limit
$q \rightarrow 1$\cite{RudGas}. In the calculation leading to a closed form
expression for the generating function of the mean field Potts model limits
were
taken in the proper order, although the final result was obtained with the
use of
non-rigorous arguments. Making the following replacements
\begin{eqnarray}
p   &=& 1 + N^{-1/3}t    ,  \label{pscaled} \\
h   &=&  HN^{-2/3}
\label{hscaled}
\end{eqnarray}
then the generating function takes the following
form \begin{eqnarray} F(p,h) &\rightarrow& \int_{-\infty}^{\infty} d \Delta
\left( \left\{ \int_0^{\infty}\exp\left[\frac{-(L-t)^3}{6} - \frac{t^3}{6} -
\Delta L \right] dL \right\} \right. \nonumber \\
&&\left. \times\Im \ln\left\{ \int_c \exp\left[ (\Delta + H)x +
\frac{x^3}{6} \right] \right\} \right) + K_c  .  \label{finitegenerating}
\end{eqnarray}
The contour integration in Eq. (\ref{finitegenerating}) is over a contour in
the complex $x$ plane that extends from $-\infty$ on the real axis to $\infty$
along a curve making an angle of 60$^{\circ}$ with respect to the positive real
axis. For details see \cite{RudGas}.

The inversion of this function
according to Eq. (\ref{inverse}) is straighforward to carry out. Shifting the
integration variable by $H$, rotating by 90$^{\circ}$ in the complex
plane,
multiplying by $e^{-imh }$ and integrating, one obtains immediately
\begin{eqnarray} n^c_m &=& N^{-2/3} \int_{\infty}^{\infty} d \Delta
\left(\exp\left[\frac{-(mN^{-2/3}-t)^3}{6} - \frac{t^3}{6} - \Delta m
N^{-2/3}\right] \right.  \nonumber \\
&& \left.\times \Im \ln \left\{\int_c \exp\left[\Delta x + \frac{x^3}{6}
\right] \right\} \right)   .  \label{finitedistribution}
\end{eqnarray}

Expression (\ref{finitedistribution}) embodies the full expected scaling form
of the cluster size distribution, and represents the mean field limit of the
distribution of cluster sizes in the case of short range bond percolation. As
such, it ought to yield the distribution of cluster sizes on a lattice in more
than six dimensions, six being the upper critical dimension for short range
bond
percolation\cite{Harris}. In addition, it constitutes the ``zeroth order''
distribution, about which one perturbs to obtain the cluster size distribution
in bond percolation in lower dimensionality.

As a test of the validity of Eq. (\ref{finitedistribution}), we have measured
the distribution of cluster sizes for mean field bond percolation on systems
with various numbers of sites, $N$. The results are displayed in Figs. 1 - 6.
The fit between the simulations and (\ref{finitedistribution}) is excellent
below the percolation transition, and when the cluster size is not too large.
When $t>0$, so that the threshold for percolation in the ``thermodynamic
limit''
has been exceeded, a feature appears in the distribution in the form of a peak
in the upper reaches of the distribution. This peak---which can be demonstrated
to have an integrated weight of unity when $t$ is large and
positive---corresponds to the contribution to the distribution of what becomes
the spanning cluster in the limit of an infinite system. As can be seen in
Figure 5, perfect agreement with simulations is not achieved for any of the
systems explored. On the other hand, there is clear evidence for convergence
between the expression (\ref{finitedistribution}) and the results of numerical
calculations as the number of sites increases to fairly large values. We are
confident that a system with the sufficient number of sites  will have a
cluster
distribution that is governed by Eq. (\ref{finitedistribution}). At this point,
we do not understand the reason for the slow approach of the data to what
appears to be its proper limiting form.

The derivation of the form (\ref{finitegenerating}) of the generating function
for percolation on a finite lattice was not entirely rigorous\cite{RudGas}. The
test of the distribution in Eq. (\ref{finitedistribution}) can thus  be regardd
as a test of that form. Given the clear evidence for agreement between the
predictions based on that form and the results of simulations, one there is
increasing confidence for the validity of the arguments that underly it.


%
%
 \begin{figure}
 \caption{The cluster size distribution, $n^c_m$, multiplied by $N^{2/3}$,
plotted against $mN^{-2/3}$, where $m$ is the size of the cluster and $N$ is
the number of sites in the system. The graph in this Figure is for $t=-1$,
where the quantity $t$ is defined in Eq. (3.1). The system is close to the
percolation transition, but the transition has not yet been reached. Note the
excellent agreement between the solid curve, representing the predictions of
Eq. (3.4) and the results of simulations for $N$ = 10,000, 40,000 and 400,000.}
 \label{1}
 \end{figure}
\begin{figure}
\caption{The cluster size distribution when $t=0$. In the bulk limit, this is
the exact location of the percolation transition.}
\label{2}
\end{figure}
\begin{figure}
\caption{The cluster size distribution when $t=1$. The system is just above the
percolation transition, and the incipient spanning cluster has begun to emerge.
The signature in the distribution function is a barely visible feature.}
\label{3}
\end{figure}
\begin{figure}
\caption{The cluster distribution when $t=2$. Now, the peak for the spanning
cluster is becoming distinct. Agreement between the analytical prediction and
the results of simulations is not nearly as good in the vicinity of this peak
as elsewhere in the Figure. However, agreement improves with increased system
size}
\label{4}
\end{figure}
\begin{figure}
\caption{The cluster distribution when $t=3$. The spanning cluster peak is well
separated from the rest of the distribution. Agreement between analysis and
simulations is not good in the vicinity of the peak, but, as previously, it
improves with increasing system size. The tendency strongly indicates
convergence.}
\label{5}
\end{figure}
\begin{figure}
\caption{a log-log plot of the distribution at the percolation transition
($t=0$). In the infinite system, this plot would have the form of a straight
line. In the finite system, a power law is obeyed until $m \propto
N^{2/3}$. This
behavior is evident in the Figure, and is displayed by both the analytical form
and the results of simulations. As in the previous Figures, the agreement
between analysis and simulations is best for the largest systems.}
\label{6}
\end{figure}
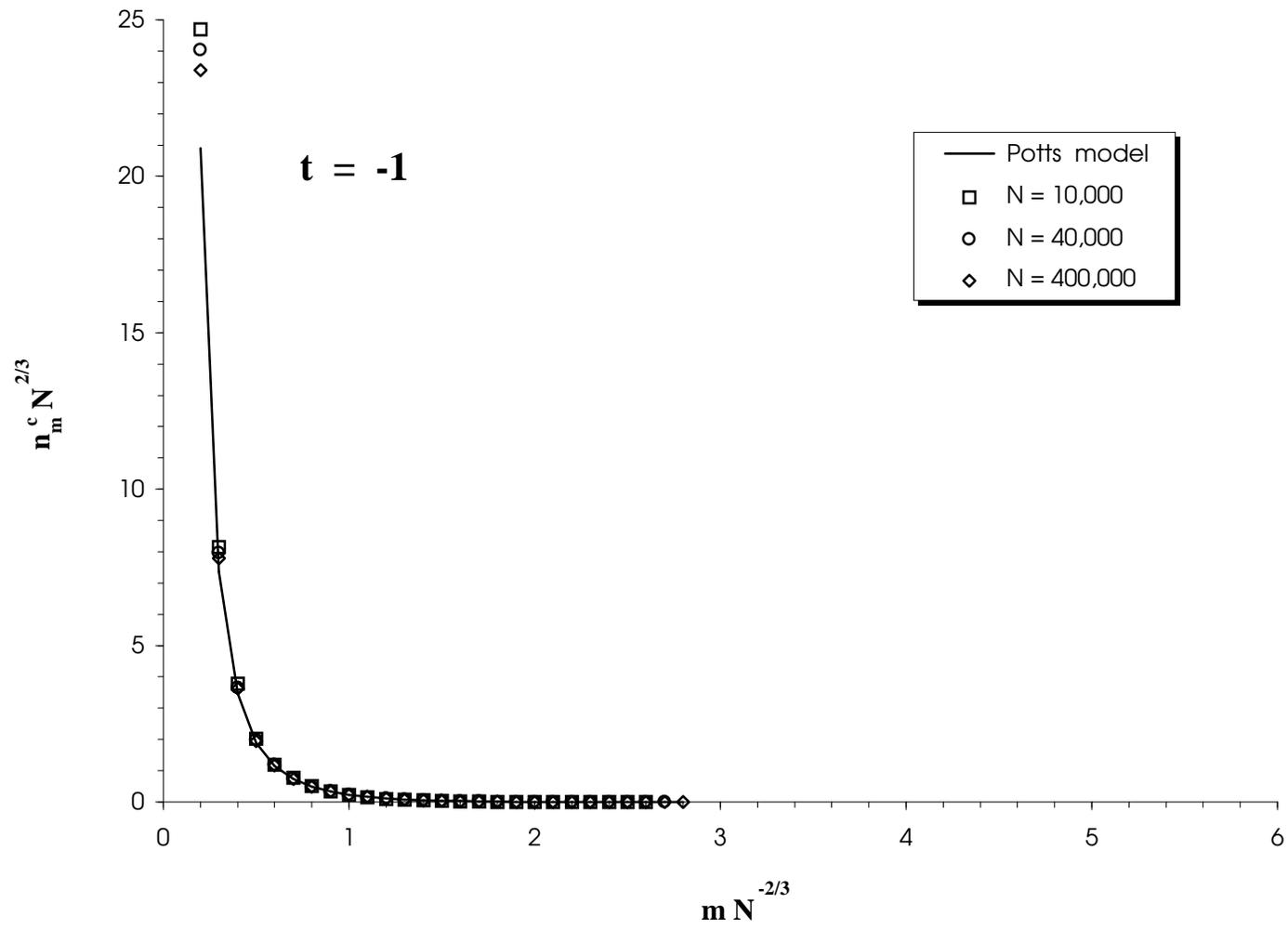
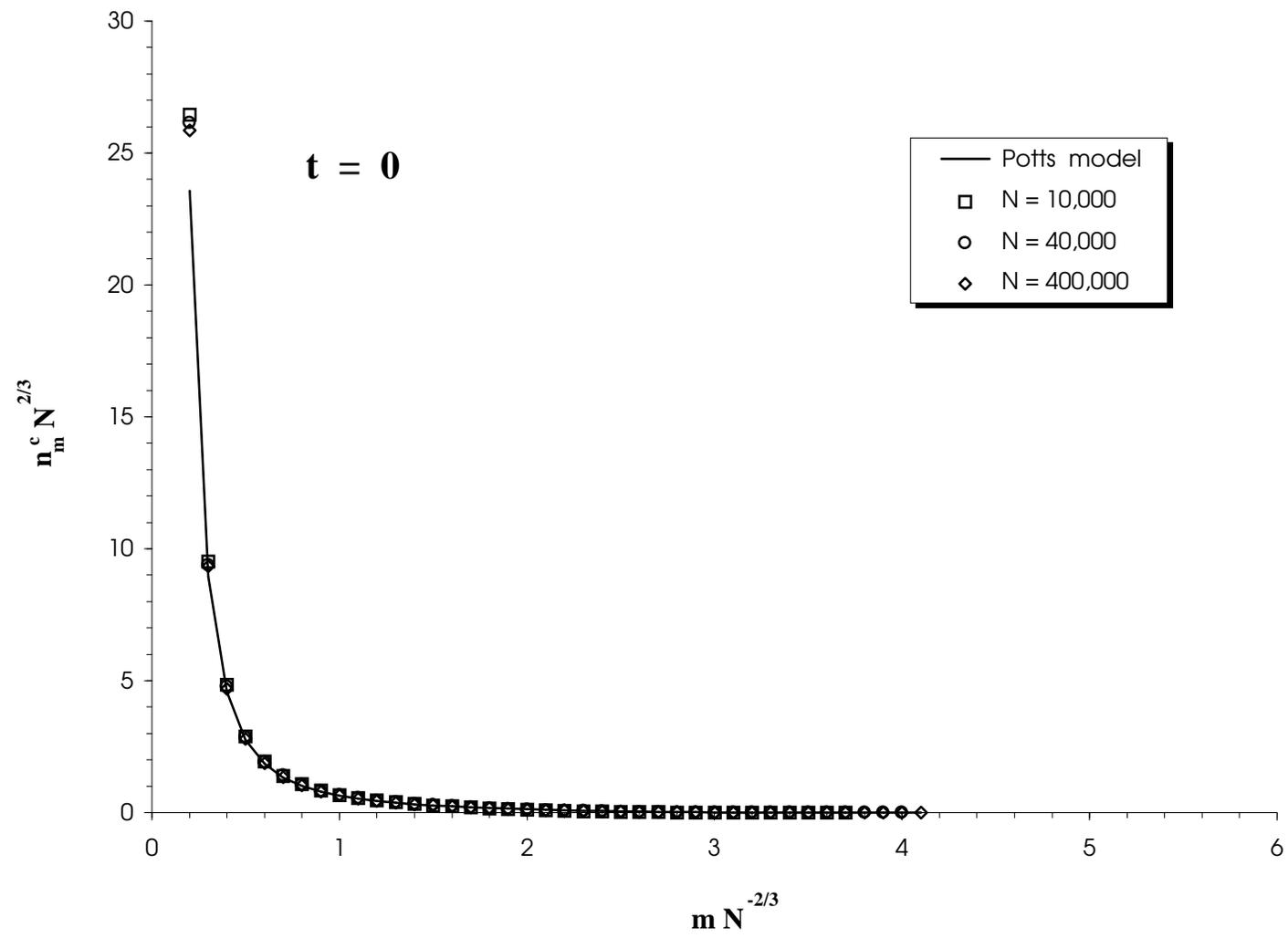
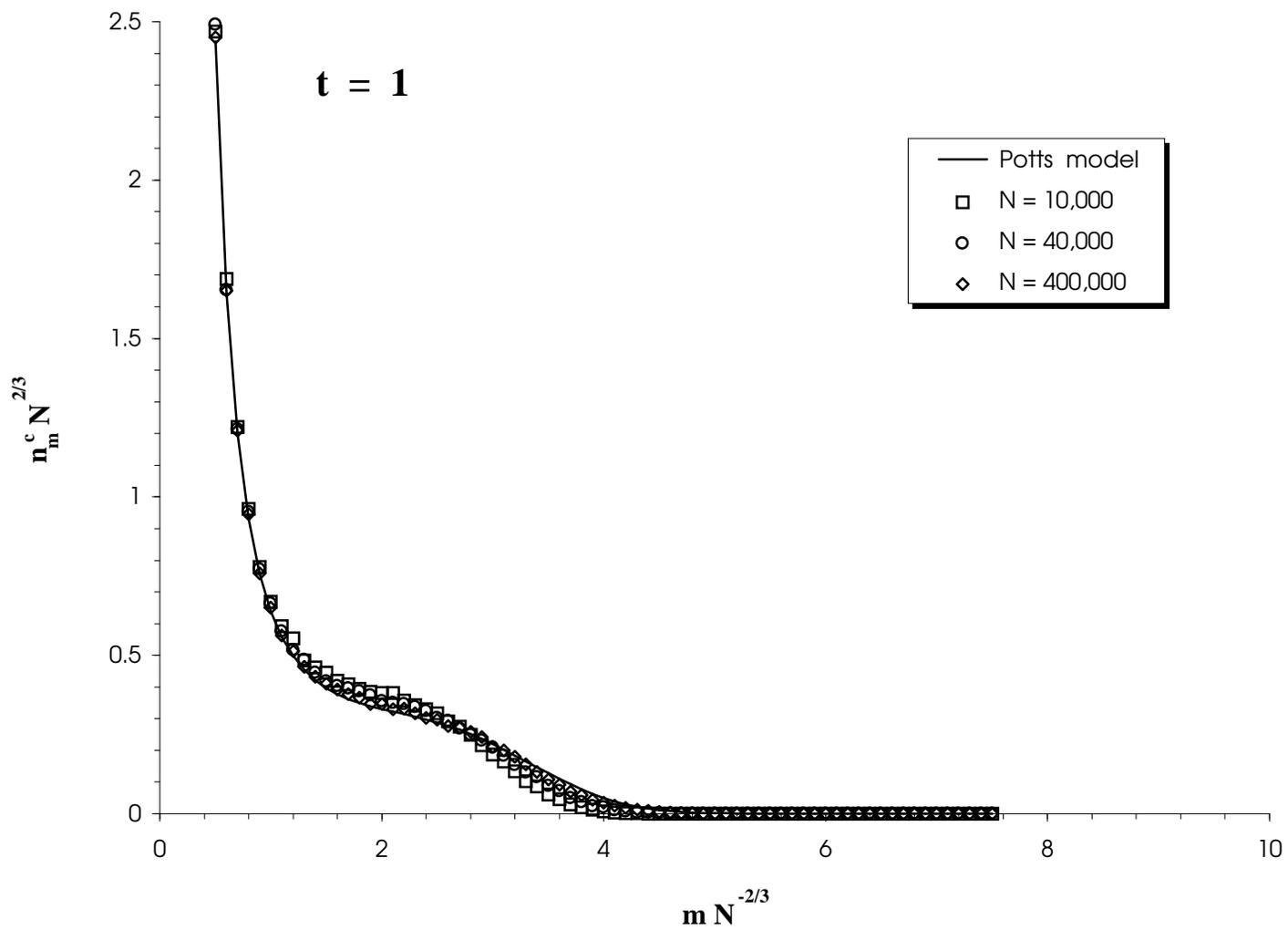
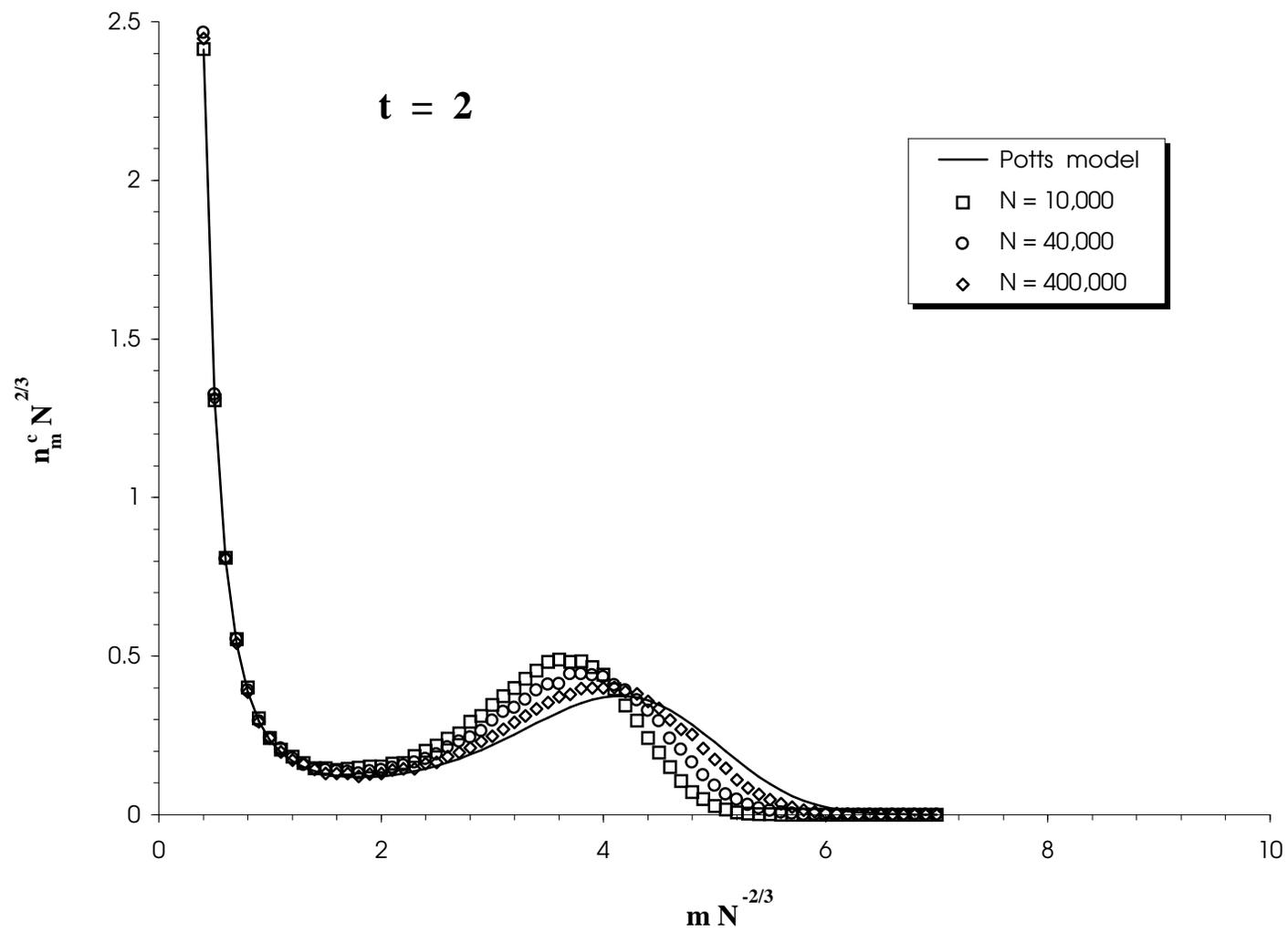
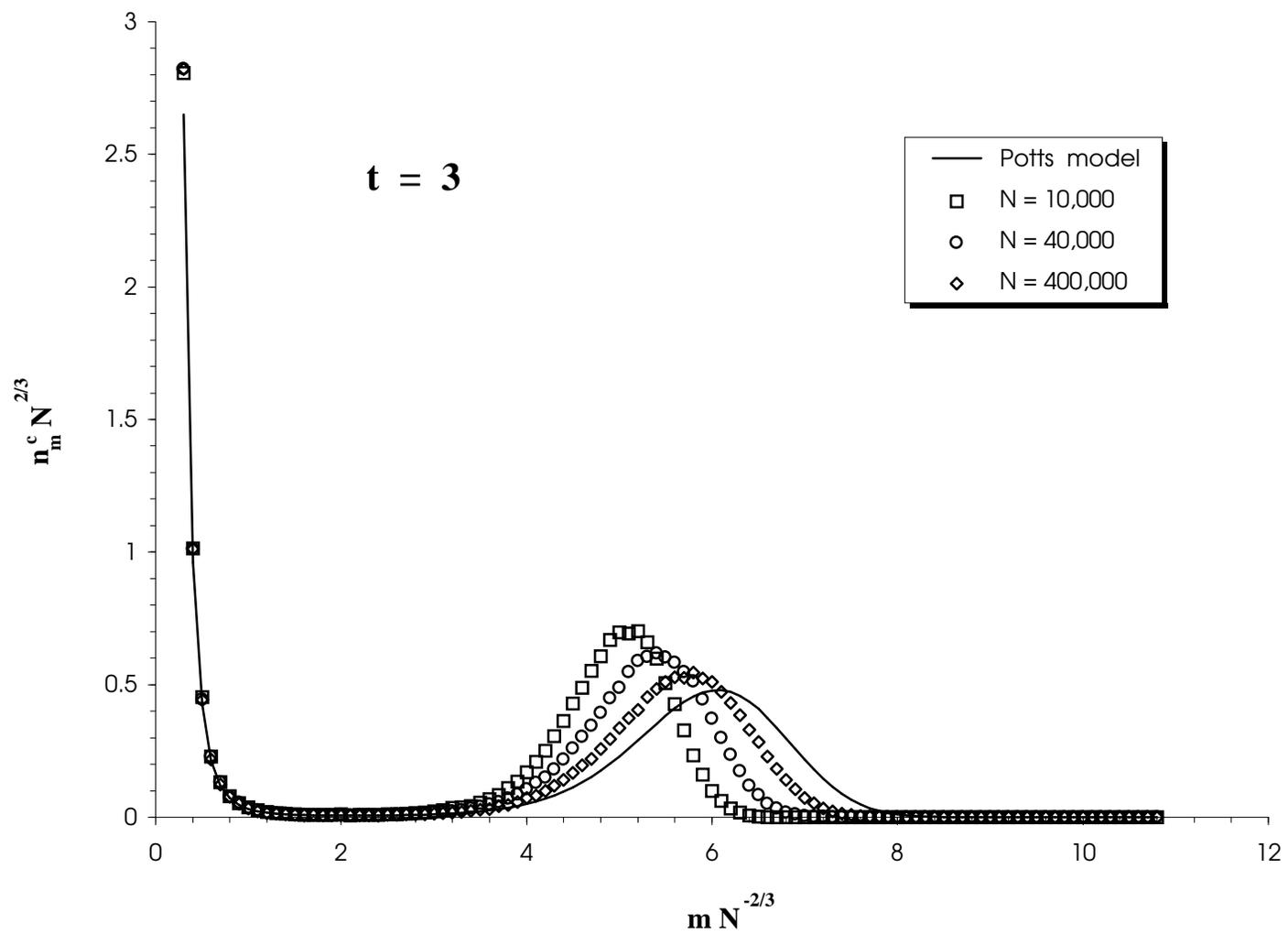
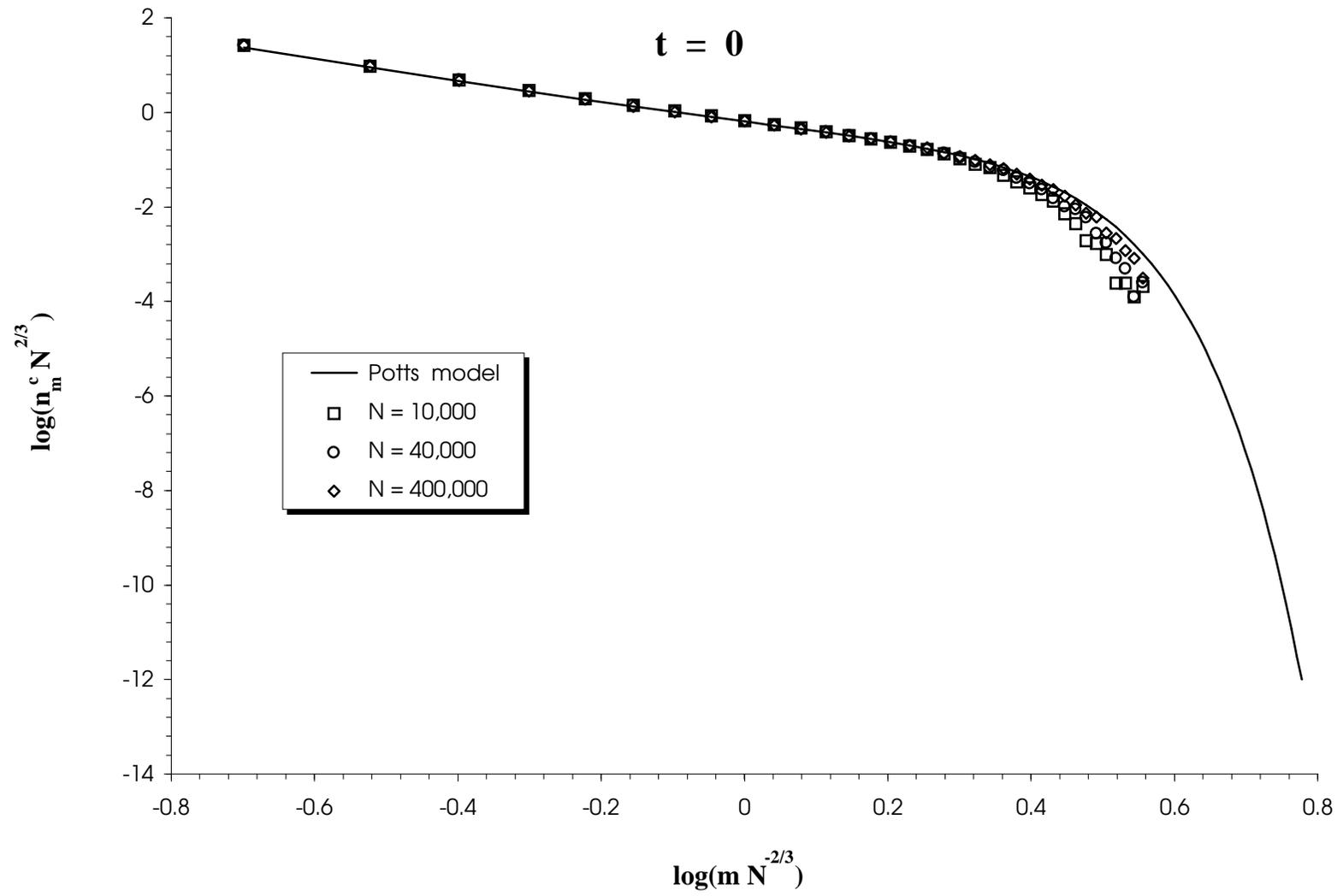
\end{document}